\documentclass[useAMS,usenatbib,usegraphicx]{mn2e}

\title[Dust emission in the far-IR as a star formation tracer at z=0: systematic trends with luminosity]{Dust emission in the far-IR as a star formation tracer at z=0: systematic trends with luminosity\thanks{Based on observations with the {\it Infrared Space Observatory (ISO)}, an ESA project with instruments funded by ESA member states (especially the PI countries: France, Germany, the Netherlands, and the United Kingdom) and with the participation of ISAS and NASA.}}
\author[D. Pierini and C. S. M\"oller]{D. Pierini$^{1}$ and C. S. M\"oller$^{2}$\\
$^{1}$Max-Planck-Institut f\"ur extraterrestrische Physik, Giessenbachstr., D-85748 Garching, Germany\\
$^{2}$Max-Planck-Institut f\"ur Astrophysik, Karl-Schwarzschild-Str. 1, D-85748 Garching, Germany}
\begin{document}

\date{Accepted ... Received ... ; in original form ...}

\pagerange{\pageref{firstpage}--\pageref{lastpage}} \pubyear{2003}

\maketitle

\label{firstpage}

\begin{abstract}
We investigate if dust emission in the far-IR continuum provides
a robust estimate of star formation rate ($SFR$)
for a nearby, normal late-type galaxy.
We focus on the ratio of the 40--1000 $\rmn{\mu m}$ luminosity
($L_{\rmn{dust}}$) to the far-UV (0.165 $\rmn{\mu m}$) luminosity,
which is connected to recent episodes of star formation.
Available total photometry at 0.165, 60, 100 and 170 $\rmn{\mu m}$
limits the statistics to 30 galaxies, which, however, span a large range
in observed (and, thus, attenuated by dust) K-band (2.2 $\rmn{\mu m}$)
luminosity, morphology and inclination ($i$).
This sample shows that the ratio of $L_{\rmn{dust}}$
to the observed far-UV luminosity depends not only on $i$, as expected,
but also on morphology and, in a tighter way, on observed K-band luminosity.
We find that
$L_{\rmn{dust}}/L_{\rmn{FUV}} \propto e^{-\tau_{\rmn{K}}~( \alpha + 0.62 )}~L_{\rmn{K}}^{0.62}$, where $L_{\rmn{FUV}}$ and $L_{\rmn{K}}$ are
the unattenuated stellar luminosities in far-UV and K, respectively,
and $\alpha$ is the ratio of the attenuation optical depths
at 0.165 $\rmn{\mu m}$ ($\tau_{\rmn{FUV}}$)
and 2.2 $\rmn{\mu m}$ ($\tau_{\rmn{K}}$).
This relation is to zeroth order independent of $i$ and morphology.
It may be further expressed as
$L_{\rmn{dust}}/L_{\rmn{FUV}} \propto L_{\rmn{K}}^{\delta}$,
where $\delta = 0.61 - 0.02 \alpha$, under the observationally-motivated
assumption that, for an average inclination,
$e^{-\tau_{\rmn{K}}} \propto L_{\rmn{K}}^{-0.02}$.
We adopt calculations of two different models of attenuation of stellar light
by internal dust to derive solid-angle averaged values of $\alpha$.
We find that $\delta$ is positive and decreases towards 0
from the more luminous to the less luminous galaxies.
This means that there is no universal ratio of far-IR luminosity
to unattenuated far-UV luminosity for nearby, normal late-type galaxies.
The far-IR luminosity systematically overestimates $SFR$
in more luminous, earlier-type spirals, owing to the increased
fractional contribution to dust heating of optical/near-IR photons
in these objects.
Conversely, it systematically underestimates $SFR$ in fainter,
later-type galaxies, whose $\tau_{\rmn{FUV}}$ is reduced.
The limited statistics and the uncertainty affecting
the previous scaling relations do not allow us to establish
quantitative conclusions, but an analogous analysis making use
of larger data-sets, available in the near future
(e.g. from {\it GALEX}, {\it ASTRO-F} and {\it SIRTF}),
and of more advanced models will allow a quantitative test of our conclusions.
\end{abstract}

\begin{keywords}
galaxies: fundamental parameters -- galaxies: ISM -- ISM: dust, extinction -- infrared: galaxies -- ultraviolet: galaxies.
\end{keywords}

\section{Introduction}

The star formation rate ($SFR$) is a fundamental property
of any galaxy at any epoch.
Star formation is a process intimately connected to
the dusty interstellar medium (ISM) of a galaxy,
at any epoch following the formation of its first-generation stars.
Attenuation of stellar radiation at different wavelengths by internal dust
is an extremely complex topic heavily discussed in the literature
(see Calzetti 2001 for a review) but its full understanding
is still far from being at reach.
Hence astronomers have tried to limit, if not overcome,
this serious problem by devising various methods for the determination
of $SFR$s, in starbursts as well as in normal (i.e., non-starburst
and non-AGN dominated) late-type galaxies (see Kennicutt 1998 for a review).
These methods, based on different astrophysical phenomena,
reach different quantitative conclusions in general
(see Charlot et al. 2002 for a discussion).
Even when the astrophysical mechanisms at the basis of these methods
produce radiation at wavelengths where dust attenuation is less effective
or not effective at all (e.g. the thermal emission from dust
at IR--mm wavelengths or the non-thermal, synchrotron emission
at radio wavelengths), there is no easy link between these radiation processes
and $SFR$ for normal late-type galaxies (e.g. Popescu et al. 2000; Bell 2002;
Pierini et al. 2003b).

The relation between dust emission in the far-IR and $SFR$
in nearby (i.e., at redshift $z = 0$) starbursts (Kennicutt 1998)
has recently been subject of particular interest, owing to the discovery
of a relevant dust-enshrouded phase of the cosmological star formation history
(see Hauser \& Dwek 2001 for a review).
The existence of an analogous relation for nearby, normal late-type galaxies
is still unclear, for the following reasons.

First, the origin of dust heating is still under debate
(see Bell 2003 for a discussion).
The pioneering work of Buat \& Xu (1996) pointed to non-ionizing,
far-UV photons as the main source of dust heating in nearby, normal
late-type galaxies.
This has been confirmed by modeling of the optical to far-IR--submm
spectral energy distributions (SEDs) available for a few such objects
(Popescu et al. 2000; Misiriotis et al. 2001).
Nevertheless the latter studies show that the contribution to dust heating
of optical and near-IR photons is not negligible, consistent with
interpretations of the mass-normalized far-IR/radio correlation
(Xu et al. 1994), and of the behaviour of the [CII](158 $\rmn{\mu m}$)
line-to-far-IR continuum emission ratio as a function of
the H$\alpha$(0.656 $\rmn{\mu m}$) equivalent width
(Pierini, Leech \& V\"olk 2003a).

Second, the standard estimate of dust emission in the far-IR,
i.e., the {\it IRAS} 40--120 $\rmn{\mu m}$ flux ``$FIR$'' (Helou et al. 1988),
may underestimate the total dust emission at far-IR--mm wavelengths
by a factor up to 3 (Popescu \& Tuffs 2002).
This is due to the fact that the {\it IRAS} spectral bands could not probe
the relevant contribution to the far-IR--mm continuum of the emission from
``cold'' (i.e., with a median temperature of 18 $\rmn{K}$) dust
(Popescu et al. 2002), rather smoothly distributed over the disk
(Haas et al. 1998; Hippelein et al. 2003).
Popescu \& Tuffs (2002) find that the mean fraction of stellar light
re-radiated by dust at 40--1000 $\rmn{\mu m}$ is about 30 per cent
among the normal Virgo Cluster late-type galaxies observed
by Tuffs et al. (2002a,b) with {\it ISOPHOT\/} (Lemke et al. 1996).
They also find that this fraction is about 15 per cent for early-type spirals
and up to about 50 per cent in some late-type disk-galaxies.

Third, dust attenuation depends on the dust/stars configuration
(i.e., geometry), structure of the dusty ISM, dust amount
and absorption/scattering properties of the mix of dust grains
(e.g. Bianchi, Ferrara \& Giovanardi 1996; Kuchinski et al. 1998;
Silva et al. 1998; Ferrara et al. 1999; Xilouris et al. 1999;
Charlot \& Fall 2000; Popescu et al. 2000; Witt \& Gordon 2000).
The grain properties may depend significantly on UV field
(Gordon, Calzetti \& Witt 1997).

Fourth, dust attenuation in the optical and near-IR seems to increase
with luminosity for nearby, normal late-type galaxies (Giovanelli et al. 1995;
Tully et al. 1998; Masters, Giovanelli \& Haynes 2003).
This is not at odds with the result of Popescu \& Tuffs (2002).
Late-type spirals and irregulars have bluer optical-to-near-IR colours
than early-type spirals (e.g. Boselli et al. 1997), which may indicate that
the former galaxies reach their maximum star formation activity more recently
than the latter ones (e.g. Kennicutt 1998; Gavazzi, Pierini \& Boselli 1996).
Thus they are expected to produce relatively more far-UV photons
than optical and, especially, near-IR ones.
However the far-UV photons are the most affected by dust attenuation
(e.g. Calzetti 2001).

The purpose of this paper is to investigate the relation between
far-IR luminosity and far-UV luminosity (produced by stars
formed in recent episodes of star formation) for nearby, normal
late-type galaxies, taking into account the observational result that
dust attenuation depends on luminosity.
We investigate how the 40--1000 $\rmn{\mu m}$-to-0.165 $\rmn{\mu m}$
luminosity ratio behaves as a function of the near-IR luminosity
(or dynamical mass -- e.g. Gavazzi et al. 1996).
This study is complementary to those of Buat \& Xu (1996), Buat et al. (1999),
Gavazzi et al. (2002) and Boselli, Gavazzi \& Sanvito (2003), who focussed on
the value of the attenuation at 0.2 $\rmn{\mu m}$ as a function of
the {\it IRAS\/} 40--120 $\rmn{\mu m}$-to-0.2 $\rmn{\mu m}$ luminosity ratio
and/or on the attenuation function\footnote{The extinction curve
describes the absorption and scattering properties of a mix of
dust grains of given size distribution and chemical composition
as a function of wavelength ($\lambda$); the attenuation function is
the convolution of the extinction curve with the geometry of
the dusty stellar system. Attenuation depends on structure of the dusty ISM
and dust amount.} ($\tau_{\rmn{\lambda}}$) for normal late-type galaxies
at $z=0$.

\section{The sample}
\begin{table*}
 \centering
 \begin{minipage}{120mm}
  \caption{Galaxy parameters.}
  \begin{tabular}{@{}llcccl@{}}
  \hline
   Denom. & Hubble type & i [deg] & $\rmn{M}_{\rmn{K}}^{\rmn{obs}}-$5log$~h$ [mag] & $\rmn{log}~\rmn{L}_{\rmn{dust}}/\rmn{L}_{\rmn{FUV}}^{\rmn{obs}}$ & Notes \\
  \hline
VCC\,66 & SBc & 71 & -21.46$\pm$0.45$^{a}$ & 0.41$\pm$0.15$^{ce}$ & HII \\
VCC\,318 & SBcd & 53 & -18.49$\pm$0.45$^{a}$ & 0.01$\pm$0.18$^{ce}$ & \\
VCC\,664 & Sc & 29 & -19.15$\pm$0.45$^{a}$ & -0.11$\pm$0.16$^{ce}$ & \\
VCC\,692 & Sc & 50 & -20.17$\pm$0.45$^{a}$ & 0.37$\pm$0.20$^{ce}$ & \\
VCC\,801 & Amorph. & 61 & -20.81$\pm$0.45$^{b}$ & 0.78$\pm$0.15$^{de}$ & HII \\
VCC\,836 & Sab & 83 & -22.33$\pm$0.45$^{a}$ & 0.83$\pm$0.17$^{ce}$ & Sy2 \\
VCC\,873 & Sc & 76 & -21.94$\pm$0.45$^{a}$ & 1.31$\pm$0.14$^{cf}$ & \\
VCC\,912 & SBbc & 50 & -20.83$\pm$0.45$^{a}$ & -0.02$\pm$0.18$^{ce}$ & \\
VCC\,1003 & S0/Sa pec & 66 & -23.72$\pm$0.45$^{a}$ & $>~$1.06$\pm$0.14$^{cf}$ & \\
VCC\,1043 & Sb (tides) & 70 & -23.47$\pm$0.45$^{a}$ & 0.76$\pm$0.20$^{ce}$ & LINER \\
VCC\,1189 & Sc & 54 & -19.00$\pm$0.45$^{a}$ & -0.01$\pm$0.18$^{ce}$ & \\
VCC\,1253 & SB0/SBa & 24 & -23.13$\pm$0.45$^{a}$ & $>~$0.33$\pm$0.14$^{cf}$ & \\
VCC\,1326 & SBa & 60 & -20.38$\pm$0.45$^{a}$ & $>~$1.13$\pm$0.14$^{cf}$ & \\
VCC\,1419 & S pec (dust) & 48 & -20.00$\pm$0.45$^{a}$ & $>~$-0.03$\pm$0.14$^{cf}$ & \\
VCC\,1450 & Sc & 31 & -19.78$\pm$0.45$^{a}$ & 0.02$\pm$0.20$^{ce}$ & \\
VCC\,1552 & Sa pec & 52 & -21.56$\pm$0.45$^{a}$ & $>~$0.65$\pm$0.14$^{cf}$ & \\
VCC\,1554 & Sm & 71 & -20.94$\pm$0.45$^{a}$ & 0.29$\pm$0.14$^{ce}$ & \\
VCC\,1678 & SBd & 24 & -18.39$\pm$0.45$^{a}$ & -0.44$\pm$0.16$^{ce}$ & HII \\
VCC\,1686 & Sm & 63 & -19.13$\pm$0.45$^{a}$ & -0.21$\pm$0.14$^{cf}$ & \\
VCC\,1690 & Sab & 65 & -23.67$\pm$0.45$^{a}$ & 0.67$\pm$0.15$^{ce}$ & LINER, Sy \\
VCC\,1699 & SBm & 50 & -18.10$\pm$0.45$^{a}$ & -0.22$\pm$0.18$^{ce}$ & \\
VCC\,1725 & Sm/BCD & 66 & -18.01$\pm$0.45$^{a}$ & -0.42$\pm$0.17$^{ce}$ & \\
VCC\,1727 & Sab & 38 & -23.95$\pm$0.45$^{a}$ & 0.72$\pm$0.18$^{ce}$ & LINER, Sy1.9 \\
VCC\,1943 & SBb & 48 & -21.50$\pm$0.45$^{b}$ & 0.18$\pm$0.16$^{de}$ & Sy1.8 \\
NGC\,3077 & I0 pec & 34 & & 0.74$\pm$0.15$^{df}$ & HII \\
NGC\,3938 & Sc & 25 & -21.83$\pm$0.45$^{b}$ & 0.42$\pm$0.18$^{df}$ & \\
NGC\,4651 & Sc & 50 & -21.40$\pm$0.45$^{b}$ & 0.13$\pm$0.15$^{de}$ & LINER \\
NGC\,4701 & Scd & 42 & & 0.26$\pm$0.18$^{df}$ & \\
NGC\,5806 & SBb & 61 & -22.23$\pm$0.45$^{b}$ & 1.10$\pm$0.18$^{df}$ & \\
NGC\,6503 & Scd & 72 & -19.77$\pm$0.45$^{b}$ & 0.58$\pm$0.14$^{df}$ & HII, LINER \\
  \hline
  \end{tabular}
 \begin{list}{}{}
  \item[$^a$] Boselli et al. (1997); $^b$ Jarrett et al. (2003); $^c$ Tuffs et al. (2002a,b); $^d$ Stickel et al. (2000);
  \item[$^e$] Brosch et al. (1997); $^f$ Rifatto et al. (1995).
 \end{list}
 \end{minipage}
\end{table*}

We select a sample of 30 nearby, normal spiral and irregular galaxies
on the basis of available {\it total\/} photometry in the far-IR -- at 60, 100
and 170 $\rmn{\mu m}$ -- (Tuffs et al. 2002a,b; Stickel et al. 2000)
and in the far-UV -- 0.165 $\rmn{\mu m}$ -- (Brosch et al. 1997;
Rifatto, Longo \& Capaccioli 1995).
In particular, 60, 100 and 170 $\rmn{\mu m}$ fluxes from Tuffs et al.
were all measured with {\it ISOPHOT\/}; conversely only the 170 $\rmn{\mu m}$
fluxes from Stickel et al. were measured with this instrument,
the 60 and 100 $\rmn{\mu m}$ fluxes listed by these authors
coming from the previous {\it IRAS\/} space mission.
All the 0.165 $\rmn{\mu m}$ measurements were taken with {\it FAUST\/}
(Lampton et al. 1993).

Furthermore, 28 galaxies out of these 30 have total photometry
in the near-IR -- $\rmn{K^{\prime}}$ band (2.1 $\rmn{\mu m}$)
or $\rmn{K_s}$ band (2.2 $\rmn{\mu m}$) -- (Boselli et al.
1997 and Jarrett et al. 2003, respectively).
We also determine the inclination ($i$) of each galaxy from its axial ratio,
listed in the RC3 (de Vaucouleurs et al. 1991), as in Bottinelli et al. (1983).

The sample under investigation contains 24 Virgo Cluster member galaxies
(Binggeli, Sandage \& Tammann 1985 -- VCC), assumed to lie at
the common distance\footnote{The Virgo Cluster is probably elongated
in the direction of the line-of-sight (e.g. Yasuda, Fukugita \& Okamura 1997).
Nevertheless, when the Tully-Fisher distances of the latter authors are used,
for 17 of the 24 VCC galaxies of this sample, the magnitude dependence in Eq. 2
of Sect. 3 does not change.} of 11.5$~h^{-1}~\rmn{Mpc}$
(as in Tuffs et al. 2002a,b), where $h$ is the Hubble constant
in units of 100 $\rmn{km~s^{-1}~Mpc^{-1}}$.
For the remaining 6 non-VCC galaxies of the sample,
we determine kinematical distances from the recession velocities
(lower than 1000 $\rmn{km~s^{-1}}$) listed in Stickel et al. (2000),
after correction to the Galactic Standard of Rest
with the NED Velocity Correction Calculator.
These 6 galaxies are not used in the determination of Eq. 2 in Sect. 3,
since their kinematical distances may be severely affected by motions
different from the Hubble free-flow.

For individual objects, the continuum emission from dust
in the 40--1000 $\rmn{\mu m}$ range comes either from Popescu et al. (2002)
or from Pierini et al. (2003b).
It is calculated through far-IR SED fitting, where each SED is fitted
with two modified-blackbody functions by constraining one parameter of the fit,
i.e., the temperature (47 $\rmn{K}$) of the ``warm'' dust emission component
(see Popescu et al. for a discussion of this method).
We assume a conservative uncertainty of 30 per cent
for these far-IR--mm fluxes.
Here we determine $L_{\rmn{dust}}/L_{\rmn{FUV}}^{\rmn{obs}}$,
i.e., the ratio of the 40--1000 $\rmn{\mu m}$ luminosity ($L_{\rmn{dust}}$)
to the observed (i.e., not corrected for internal dust attenuation
but corrected for Galactic extinction as in Burstein \& Heiles 1982)
stellar luminosity at 0.165 $\rmn{\mu m}$ ($L_{\rmn{FUV}}^{\rmn{obs}}$).
Errors in the observed 0.165 $\rmn{\mu m}$ fluxes are listed
in the data sources.

Furthermore, we assume that $\rmn{K^{\prime} - K} = 0$ or $\rmn{K_s - K} = 0$
for all the sample galaxies.
Thus we will speak of K-band magnitudes (or luminosities) hereafter.
We assume an error of 0.45 mag for the observed (i.e, not corrected for
internal dust attenuation but corrected for the negligible Galactic extinction
in K) K-band absolute magnitude ($M_{\rmn{K}}^{\rmn{obs}}$) of each galaxy,
which is comprised of an error of 0.1 mag in the photometry
and an error of 20 per cent in the distance.

Tab. 1 lists the galaxy parameters relevant to this study.

\section{Results}
\begin{figure}
 \includegraphics[width=84mm]{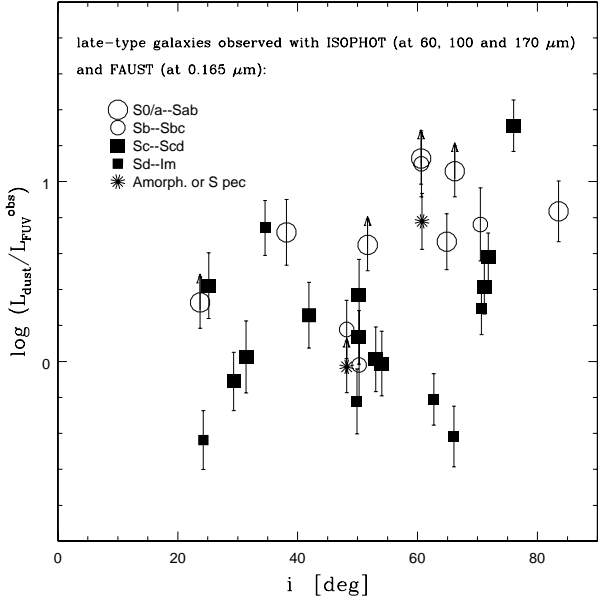}
 \caption{The 40--1000 $\rmn{\mu m}$-to-observed 0.165 $\rmn{\mu m}$
luminosity ratio ($L_{\rmn{dust}}/L_{\rmn{FUV}}^{\rmn{obs}}$) as a function of
galaxy inclination ($i$). Different symbols reproduce the 30 sample galaxies
according to their morphology. Arrows identify galaxies with upper limits
to $L_{\rmn{FUV}}^{\rmn{obs}}$. $L_{\rmn{dust}}/L_{\rmn{FUV}}^{\rmn{obs}}$
depends on $i$ and on Hubble type.}
 \label{Fig.1}
\end{figure}
\begin{figure}
 \includegraphics[width=84mm]{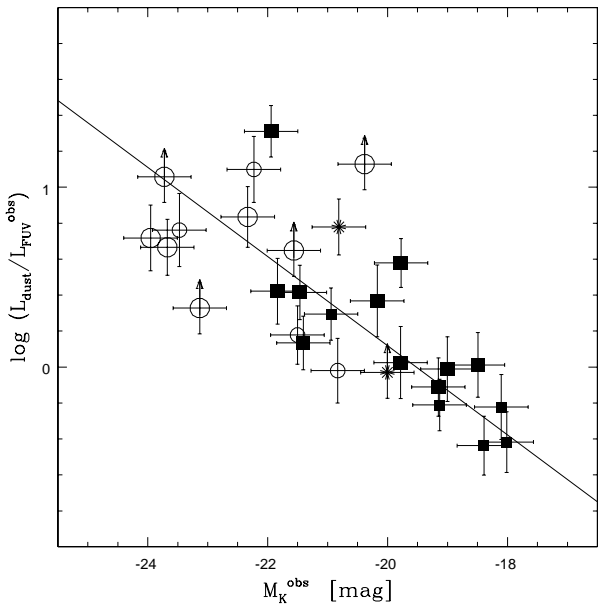}
 \caption{$L_{\rmn{dust}}/L_{\rmn{FUV}}^{\rmn{obs}}$ as a function of
the absolute K-band magnitude, not corrected for internal dust attenuation,
($M_{\rmn{K}}^{\rmn{obs}}$) for 28 galaxies with K-band data out of
the 30 galaxies plotted in Fig. 1. The solid line reproduces
the best linear-fit to the data of 19 VCC galaxies detected
at 0.165 $\rmn{\mu m}$.}
 \label{Fig.2}
\end{figure}

Fig. 1 reproduces the 40--1000 $\rmn{\mu m}$-to-observed 0.165 $\rmn{\mu m}$
luminosity ratio $L_{\rmn{dust}}/L_{\rmn{FUV}}^{\rmn{obs}}$
as a function of galaxy inclination $i$.
Galaxies are grouped in 5 classes: S0/a--Sab, Sb--Sbc, Sc--Scd, Sd--Im,
and amorphous or peculiar spirals.
From the visual inspection of Fig. 1, we conclude that:
\begin{enumerate}
  \item $L_{\rmn{dust}}/L_{\rmn{FUV}}^{\rmn{obs}}$ tends to increase
with increasing galaxy inclination, whatever the galaxy morphology;
  \item early-type spirals tend to have larger values of
$L_{\rmn{dust}}/L_{\rmn{FUV}}^{\rmn{obs}}$ than later ones\footnote{The only
exception to this picture is NGC\,3077. We have no obvious explanation
for that.}.
\end{enumerate}
The first trend is easy to interpret:
\begin{equation}
   L_{\rmn{FUV}}^{\rmn{obs}} = L_{\rmn{FUV}}~e^{-\tau_{\rmn{FUV}}},
\end{equation}
where $\tau_{\rmn{FUV}}$ and $L_{\rmn{FUV}}$ are the attenuation optical depth
and the unattenuated luminosity at 0.165 $\rm{\mu m}$, respectively.
Light in the far-UV is produced by young stars associated with the galaxy disk.
Dust is observed to be distributed only in the disk of a late-type galaxy,
whatever the Hubble type.
Thus $\tau_{\rmn{FUV}}$ increases with $i$ owing to the increasing
disk-averaged dust column density along the line-of-sight,
for fixed structure of the ISM, dust amount and extinction curve.

The second trend in Fig. 1 is new and more interesting.
It suggests that $L_{\rmn{dust}}/L_{\rmn{FUV}}^{\rmn{obs}}$
depends not only on dust properties (geometry, distribution, clumpiness,
total amount and grain type) but also on stellar properties.

Both stellar properties and dust attenuation seem to be related
to near-IR luminosity (cf. Introduction).
For this reason, in Fig. 2, we plot $L_{\rmn{dust}}/L_{\rmn{FUV}}^{\rmn{obs}}$
vs. $M_{\rmn K}^{\rmn{obs}}$, the absolute K-band magnitude
not corrected for internal dust attenuation.
Fig. 2 shows that $L_{\rmn{dust}}/L_{\rmn{FUV}}^{\rmn{obs}}$
is larger for more luminous galaxies, whatever their morphology.

A least-squares linear fit to the data of the 19 VCC galaxies of Tab. 1
with K-band data and detected at 0.165 $\rmn{\mu m}$ gives:
\begin{equation}
   log (L_{\rmn{dust}}/L_{\rmn{FUV}}^{\rmn{obs}}) = -0.25 (\pm 0.03) M_{\rmn{K}}^{\rmn{obs}} - 4.84 (\pm 0.55)
\end{equation}
with a reduced $\chi^2$ equal to 2.3.
This fit well describes the behaviour of all galaxies in Fig. 2.
It is to zeroth order independent of $i$ and Hubble type.
Eq. 2 means that
\begin{equation}
   L_{\rmn{dust}}/L_{\rmn{FUV}} \propto e^{-\tau_{\rmn{K}}~( \alpha + 0.62 )}~L_{\rmn{K}}^{0.62},
\end{equation}
where $\tau_{\rmn{K}}$ and $L_{\rmn{K}}$ are
the K-band attenuation optical depth and unattenuated luminosity, respectively,
while $\alpha$ represents the far-UV--K slope of the attenuation curve, being
\begin{equation}
   \alpha = \tau_{\rmn{FUV}} / \tau_{\rmn{K}}.
\end{equation}

Both $\tau_{\rmn{K}}$ and $\alpha$ are expected to depend on $L_{\rmn{K}}$
in a non-straightforward way, as we will discuss in Sect. 4.
However observations show that the inclination-averaged attenuation
optical depth (e.g. $\tau_{\rmn{K}}$) increases with luminosity
(cf. Introduction).
Conversely, neither the behaviour of $\alpha$ with luminosity
has been established observationally, nor the dependence of $\alpha$
on dust attenuation has been thoroughly investigated through models
of radiative transfer yet.
In Sect. 4, we will explore how the solid-angle averaged $\alpha$
depends on the solid-angle averaged attenuation optical depth in K band,
whatever the galaxy morphology, through two different models of attenuation
of stellar light by internal dust.
Here we determine the dependence of $\tau_{\rmn{K}}$ on $L_{\rmn{K}}$,
averaged over $i$ and Hubble type.

We extract this information from the sample of Masters et al. (2003),
since large statistics are needed for a robust conclusion.
By doing so, we may introduce a systematic bias in the following analysis,
which arises from the different selection criteria of the Masters et al.
sample and the subsample of 19 VCC galaxies used in the determination of Eq. 2.
We will discuss this issue in Sect. 4.

For a sample of 15,224 nearby spiral galaxies of different morphology,
with known $z$, and photometry from the 2 Micron All-Sky Survey (2MASS)
Extended Source Catalog (Jarrett et al. 2003), Masters et al. (2003) find that
the correction to face on of $M_{\rmn{K}}^{\rmn{obs}}$ can not be described
by a simple linear dependence on the decimal logarithm
of the major-to-minor axis ratio ($log (a/b)$).
Furthermore, they find a bi-linear dependence of
the correction to face on of $M_{\rmn{K}}^{\rmn{obs}}$
on $M_{\rmn{K}}^{\rmn{obs}}$ itself, when they bin their sample
in luminosity and, thus, take an average over inclination
and Hubble type in each bin (their Fig. 13)\footnote{This result is obtained
under the assumption that all disk-galaxies are transparent
at the 21 K-mag arcsec$^{-2}$ isophotal radius (see Masters et al. 2003
for a discussion).}.
In particular, Masters et al. conclude that the average K-band attenuation
is null for galaxies with $M_{\rmn{K}}^{\rmn{obs}}$ fainter than -20.9,
it increases with increasing luminosity for galaxies with
$-22.2 < M_{\rmn{K}}^{\rmn{obs}} \le -20.9$, it is constant for galaxies
with $M_{\rmn{K}}^{\rmn{obs}} \le -22.2$.
However, Masters et al. note that a linear fit to their data
is not inconsistent with the data errors.
A linear trend is consistent with the results obtained by
Giovanelli et al. (1995), in a way analogous to Masters et al.
but for a sample of 2816 Sb--Sd (mostly Sc) galaxies observed in I band
(0.81 $\rmn{\mu m}$), and, independently, by Tully et al. (1998),
for a sample of 87 spiral galaxies observed in the optical and near-IR.

Hence we assume that the inclination-averaged K-band attenuation
changes smoothly with $L_{\rmn{K}}$, whatever the galaxy morphology,
and, in particular, that
\begin{equation}
   e^{-\tau_{\rmn{K}}} \propto L_{\rmn{K}}^{- \beta},
\end{equation}
where $\beta > 0$.
We determine $\beta$ from Fig. 13 of Masters et al. (2003) as follows:
\begin{equation}
   M_{\rmn{K}}^{\rmn{obs}} - M_{\rmn{K,0}} = ( -0.04~M_{\rmn{K}}^{\rmn{obs}}-0.75 )~log (a/b),
\end{equation}
where $M_{\rmn{K,0}}$ is the absolute K-band magnitude corrected
for attenuation to face on, and $log (a/b)$ is the decimal logarithm
of the major-to-minor axis ratio, averaged over $i$ and Hubble type.
Eq. 6 can be re-expressed as:
\begin{equation}
   e^{-\tau_{\rmn{K}}} \propto e^{-\tau_{\rmn{K,0}}} (L_{\rmn{K}}~e^{-\tau_{\rmn{K}}})^{-0.04~log (a/b)},
\end{equation}
where $\tau_{\rmn{K,0}}$ is the attenuation optical depth in K band
for a face-on disk galaxy.
$\tau_{\rmn{K,0}}$ depends on $L_{\rmn{K}}$, of course,
but is reasonably expected to be very close to 0 for all luminosities,
from the results of Masters et al., so that it introduces
a second order effect in Eq. 7.
Thus we substitute $e^{-\tau_{\rmn{K,0}}}$ with 1 in Eq. 7.
In addition, we determine an inclination-averaged value for $log (a/b) = 0.48$,
considering that the intrinsic spatial thickness of the stellar disk
is a function of the Hubble type of its parent galaxy (Bottinelli et al. 1983).
After substitution of the latter value in Eq. 7, we obtain a relation
between the inclination-averaged K-band attenuation and $L_{\rmn{K}}$,
where $\beta \sim 0.02$.
Hence Eq. 3 becomes:
\begin{equation}
   L_{\rmn{dust}}/L_{\rmn{FUV}} \propto L_{\rmn{K}}^{\delta},
\end{equation}
where $\delta = 0.61 - 0.02 \alpha$.
This is our main observational result.

\section{Discussion and conclusions}

Eq. 8 derives from our data (Eq. 3) under the assumption that
the inclination-averaged K-band attenuation changes smoothly with $L_{\rmn{K}}$
(Eq. 5), whatever the galaxy morphology.
We calibrate the latter behaviour with the data of Masters et al. (2003)
(Eq. 6).
We adopt linear fits to our data (Fig. 2) and to those of Masters et al.
(their Fig. 13) for the only reason that a linear dependence
is the simplest way to interpret the behaviour of the quantities
in each of these two figures.
With this caveat in mind, we recall that $\tau_{\rmn{K}}$ and $\alpha$ in Eq. 3
are expected to depend on $L_{\rmn{K}}$ for at least two reasons.

First, the disk-averaged dust column density may be expressed as
the disk-averaged product of metal abundance (relative to H),
hydrogen-column density and depletion factor of metals onto dust grains.
Disk metallicity increases with luminosity
(Zaritsky, Kennicutt \& Huchra 1994) but the mass-fractions of
atomic, molecular and total (i.e., atomic$+$molecular) hydrogen
seem to decrease with luminosity (Gavazzi et al. 1996).

Second, in opposition to the far-UV luminosity, the K-band luminosity
is produced also by old stars, part of them residing also in the central,
ellipsoidal bulge, if any.
Intuitively, the difference in dust/stars distribution between bulge and disk
components of a bulge$+$disk system is expected to produce
different attenuation curves for these two components,
for fixed properties of the dusty ISM and galaxy inclination.
Everything else being fixed, the attenuation curve of a bulge$+$disk system
will depend on the $\lambda$-dependent bulge-to-disk luminosity ratio.
Thus, $\tau_{\rmn{K}}$ and $\alpha$ will depend on the bulge-to-disk
K-band luminosity ratio and, thus, on K-band luminosity,
since this ratio increases with K-band luminosity (e.g. Graham 2001).

Observations show that dust attenuation in the optical and near-IR
increases with increasing luminosity for nearby normal late-type galaxies,
whatever the Hubble type (Giovanelli et al. 1995; Tully et al. 1998;
Masters et al. 2003).

In Sect. 3, we determine this dependence from the data of Masters et al.,
whose sample comprises 15,244 nearby, normal spiral galaxies
with $c z > 3000~\rmn{km~s^{-1}}$ (the redshift $z$ is determined from
the Doppler effect in the HI line), selected from the Arecibo General Catalog
(a private compilation by these authors known as the AGC),
and with available near-IR photometry (Jarrett et al. 2003).
The luminosity distribution of the Masters et al. sample peaks
at $M_{\rmn{K}}^{\rmn{obs}} \sim -23$, most of the galaxies of their sample
being within one magnitude of this (see Fig. 11 in Masters et al.).
Conversely, the subsample of 19 VCC galaxies used to determine Eq. 2
represents equally well late-type galaxies
with $M_{\rmn{K}}^{\rmn{obs}} \le -18$ (cf. Tab. 1 and Fig. 2).
The overlap in K-band luminosity of these two samples
is far from being slim: the Masters et al. sample extends
to K-band magnitudes as faint as ours (their Fig. 11).
The fraction of their galaxies fainter than -20.6 in K is a few per cent
of those at the peak of the K-band luminosity distribution,
but it still corresponds to several hundreds of galaxies.
Nevertheless, the faintest-luminosity bin in Fig. 13 of Masters et al.
is dominated by objects with $-21.2 \le M_{\rmn{K}}^{\rmn{obs}} \le -20$.
Hence, the adoption of Eq. 6 for the subsample of 19 VCC galaxies
used to determine Eq. 2 is an extrapolation of the Masters et al. results
to K-band magnitudes as faint as -18.

The incompleteness of the Masters et al. sample at K-band magnitudes
fainter than -23 may be related primarily to the sensitivity limit of
radio telescopes in the HI line and, secondarily,
to the limiting K-band magnitude (13.4 mag) of the 2MASS (Jarrett et al. 2000).
Detection in the HI line surely depends on amount and distribution of HI gas
and, thus, may be expected to bias the resultant sample against ``HI anaemic''
or, in general, HI-poor spirals, whatever the galaxy luminosity and morphology.
On the other hand, selection in K band bias the resultant sample
against late-type, gas-rich galaxies, since the K-band surface brightness
decreases towards later-type galaxies (e.g. Jarrett et al. 2003).
Conversely, Eq. 2 derives from 19 galaxies mostly selected from
the optically complete sample observed by Tuffs et al. (2002a,b)
with {\it ISOPHOT}, according to the availability of detections
in the far-IR and far-UV (cf. Sect. 2).
Given the complex relations between metal abundance, HI-gas content
and luminosity, it is difficult to establish whether
the Masters et al. (2003) sample is indeed biased towards higher
or lower internal extinction with respect to the subsample of Eq. 2.
Thus we acknowledge the uncertain derivation of the relation between
dust attenuation and luminosity (Eq. 6) (see also discussion in
Masters et al.) and its extension to the subsample of Eq. 2.
However, we anticipate that the value of $\delta$ (Eq. 8)
at the high-luminosity end of the galaxy K-band luminosity function
is robust against an error in $\beta$ (Eq. 5) of a factor of 2 or so.

At this point, we note that the interpretation of our main observational result
(Eq. 8) depends critically on the value of the parameter $\delta$,
i.e. on the far-UV--K slope of the attenuation function (Eq. 4),
averaged over galaxy inclination and Hubble type.
We take this information from existing models of attenuation
of stellar light by internal dust.
In particular, we consider those of Witt \& Gordon (2000)
and Charlot \& Fall (2000), because these two investigations focus on
different aspects of internal-dust attenuation in galaxies.

The Monte Carlo simulations of multiple-scattering radiative transfer
by Witt \& Gordon discuss in detail the dependence of $\tau_{\rmn{\lambda}}$
on geometry, structure of the dusty ISM (homogeneous vs. two-phase clumpy),
dust column density
(parameterized by $N_{\rmn{dust}}$\footnote{$N_{\rmn{dust}}$ is the value
of the radial attenuation optical depth at V band (0.55 $\rmn{\mu m}$),
from the center to the edge of the dust environment, where dust has
a constant density, homogeneous distribution. It gives the amount of dust
of each Witt \& Gordon model.}), and grain properties
(Milky Way-type dust vs. Small Magellanic Cloud-type dust).
Witt \& Gordon assume three different, spherical dust/stars configurations,
ranging from a geometry (DUSTY) where dust and stars are intermixed,
to a geometry (SHELL) where dust and stars are separated.
We calculate values of $\alpha$ from the attenuation functions
listed by Witt \& Gordon, for the range 0.25--8 in $N_{\rmn{dust}}$.
This range provides values of $\tau_{\rmn{K}}$ between 0 and 0.11 (or larger),
where 0.11 is the range in $\tau_{\rmn{K}}$ expected from Eq. 5,
for the range of 2.4 orders of magnitude in $L_{\rmn{K}}$ (Fig. 2).
Apart from the homogeneous SHELL models, which are anyway very far from
describing a real dusty galaxy (e.g. Calzetti 2001), for all other models
$\alpha$ ranges between 2.8 and 40.7, $\alpha$ increasing by almost
a factor of 5, on average, when $N_{\rmn{dust}}$ decreases by a factor of 32.
This increase in $\alpha$ depends not only on geometry but also on dust-type,
since the SMC extinction curve has a larger (smaller) value
at 0.165 (2.2) $\rmn{\mu m}$ than the MW extinction curve,
and on structure of the ISM, since a clumpy medium is more transparent
than a homogeneous one.
From the previous models we conclude that the parameter $\delta$ (Eq. 8)
is positive, in general, and decreases towards 0 when going from
the more luminous (i.e., the more attenuated -- cf. Eq. 5) galaxies
of our sample to the less luminous (i.e., the less attenuated) ones.
Slightly negative values of $\delta$ may be reached,
if SMC-type dust is present.

We note that, in the Witt \& Gordon models, the dust/stars distribution
is independent of the characteristic age of the stellar population.
However, younger stellar populations are associated with dustier regions,
and vice versa for the older ones (e.g. Kuchinski et al. 1998;
Silva et al. 1998; Xilouris et al. 1999; Charlot \& Fall 2000;
Popescu et al. 2000).
For this reason the previous estimates of $\alpha$ and $\delta$
must be considered ``cum grano salis''.
In particular, they constitute lower and upper limits, respectively.

The results of Fig. 2 can also be interpreted in the framework of
the Charlot \& Fall (2000) models, which describe the effects of dust on
the integrated spectral properties of galaxies, based on an idealized
description of the ISM.
One of the key features of these models is the accounting of
the different attenuations affecting young and old stars in galaxies,
as manifested through the different attenuation of line and continuum photons
(Calzetti 2001).
Charlot \& Fall show that the observed ultraviolet, infrared,
H$\alpha$ and H$\beta$(0.486 $\rmn{\mu m}$) luminosities
of nearby starburst galaxies constrain the wavelength dependence of primarily
the attenuation affecting old stars in galaxies, as young stars
are typically embedded in optically thick, giant molecular clouds.
The stellar population-averaged attenuation curve depends on
the relative ratio of young and old stars in a galaxy,
and hence on the star formation history (see their Fig. 5).
For the purpose of this study, we compute Charlot \& Fall models
with an exponentially declining star formation history of
characteristic timescale 6 Gyr (typical of a spiral galaxy).
For the standard case in which the attenuation optical depth
affecting stars older than $10^7$ yr is one third of
the attenuation optical depth at V band $\tau_{\rmn{V}}$ seen
by young stars, we find that, for $\tau_{\rmn{V}}=0.3$ (8.0),
$\alpha$ ranges from 10.6 to 12.9 (6.3 to 6.9) at ages between 0.3 and 6.0 Gyr.
If the ratio of the attenuation optical depths affecting old and young stars
is lowered from 0.3 to 0.1, $\alpha$ is found to range from 16.6 to 29.5
(6.4 to 8.5).
Overall, this implies $\delta \sim 0.02$--0.48, with $\delta$ increasing
with decreasing $\tau_{\rmn{V}}$ at fixed age.

At this point the reader may wonder how a change of the bulge-to-disk
luminosity ratio affects the solid-angle averaged values
of $\alpha$ and $\delta$.
Honestly, it is hard to predict the results of radiative transfer models
for a bulge$+$disk geometry.
Intuitively, however, we may expect that the inclination-averaged
attenuation optical depth at optical/near-IR wavelengths is larger
for the bulge than for the disk, since, for most of the possible values of $i$,
roughly half of the bulge light will be affected by
the dustiest, central region of the dusty double-exponential disk,
while the disk light is affected by the whole, extended dust distribution.
We may use the Charlot \& Fall models to describe this effect.
An increasing bulge-to-disk luminosity ratio would imply that
the attenuation optical depth of the old stars increases
with respect to the attenuation optical depth of the stars younger
than $10^7$ years.
As shown before, an increase of this ratio (e.g. from 0.1 to 0.3)
would decrease $\alpha$.
As a consequence, $\delta$ is positive over a wide range of luminosities
and Hubble types and increases with increasing K-band luminosity\footnote{In
the Charlot \& Fall models, the attenuation function may be constrained
from observations (or even determined in a probabilistic way),
without specifying the spatial distribution and optical properties of the dust.
Other models of attenuation of stellar light by internal dust
include the 3D distribution of the stars, in bulge (if any) and disk,
and of the dust, in the disk (e.g. Xilouris et al. 1999; Popescu et al. 2000).
It is beyond the goals of this paper to compare values
of the solid-angle averaged $\alpha$ obtained from different models.
Anyway, we do not expect eventual differences to erase our conclusion
on the parameter $\delta$, since all the models of dust attenuation predict
that $\tau_{\rmn{\lambda}}$ becomes ``greyer'' (i.e., that $\alpha$ decreases)
when the dust amount increases.}.

The main uncertainty in Eq. 8 is the sign of the parameter $\delta$
at the very-faint end of the galaxy K-band luminosity function.
In fact, both the Witt \& Gordon (2000) and the Charlot \& Fall (2000) models
predict large values of $\alpha$ for very low attenuations
(i.e., very low luminosities).
Hence the precise value of $\delta$ for the least luminous galaxies
depends critically on the statistical behaviours determined from Fig. 2
of this work (Eq. 2), and from Fig. 13 of Masters et al. (2003) (Eq. 6).

To summarise, $L_{\rmn{dust}}/L_{\rmn{FUV}}$ and $L_{\rmn{K}}$ seem to increase
together for a wide range in dust and stellar population parameters.
Thus our analysis points toward the absence of a universal ratio
of far-IR luminosity to unattenuated far-UV luminosity
for normal late-type galaxies at $z=0$.

The relative contribution of the emission in the far-UV
to the bolometric stellar luminosity of a galaxy is expected to decrease
with increasing K-band luminosity (or dynamical mass -- Gavazzi et al. 1996)
or towards earlier spirals (Kennicutt 1998).
In this case, the larger fraction of optical/near-IR photons produced
by the stellar populations and the larger attenuation at these wavelengths
positively couple together so that the relative contribution of
the far-UV photons to dust heating decreases.

Thus the far-IR luminosity systematically overestimates $SFR$
in more luminous, earlier spiral galaxies, where a larger fraction of
the dust heating is due to optical/near-IR photons, emitted by
intermediate/low-mass stars formed mostly in relatively old episodes
of star formation.
Conversely, it systematically underestimates $SFR$ in less luminous,
later spiral/irregular galaxies, where the maximum star formation activity
is more recent but the far-UV attenuation optical depth is lower.

The limited statistics and the uncertainty affecting the scaling relations
that we determine (Eq. 2 and 6) do not allow us to establish
quantitative conclusions, but an analogous analysis making use
of larger data-sets, available in the near future
(e.g. from {\it GALEX}, {\it ASTRO-F} and {\it SIRTF}),
and of more advanced models will allow a quantitative test of our conclusions.

\section*{Acknowledgments}

We are grateful to S. Charlot for useful discussions and his comments
on this manuscript.
D.P. also acknowledges useful discussions with A.N. Witt.
We thank the referee, C.C. Popescu, for her detailed, insightful comments
and suggestions.
C.M. thanks the Alexander von Humboldt Foundation, the Federal Ministry
of Education and Research, and the Programme for Investment in the Future (ZIP)
of the German Government for their support.
This research has made use of the NASA/IPAC Extragalactic Database (NED)
which is operated by the Jet Propulsion Laboratory, California Institute
of Technology, under contract with the National Aeronautics
and Space Administration.


\bsp

\label{lastpage}

\end{document}